\documentclass[conference]{IEEEtran}
\IEEEoverridecommandlockouts

\usepackage{cite}
\usepackage{bm}
\usepackage{amsmath,amssymb,amsfonts}
\usepackage{algorithm}
\usepackage{algorithmic}
\usepackage{graphicx}

\usepackage{graphicx}
\usepackage{subcaption}

\usepackage[letterpaper, left=0.625in, right=0.625in, bottom=1.02in, top=0.70in]{geometry}

\usepackage{textcomp}
\usepackage{xcolor}

\newcommand{\be}{\begin{equation}}
\newcommand{\ee}{\end{equation}}

\def\BibTeX{{\rm B\kern-.05em{\sc i\kern-.025em b}\kern-.08em
    T\kern-.1667em\lower.7ex\hbox{E}\kern-.125emX}}

\begin{document}

\title{HawkRover: An Autonomous mmWave Vehicular Communication Testbed with Multi-sensor Fusion and Deep Learning
}
\author{\IEEEauthorblockN{Ethan Zhu\IEEEauthorrefmark{1}, Haijian Sun\IEEEauthorrefmark{2}, Mingyue Ji\IEEEauthorrefmark{3}} \\
\IEEEauthorblockA{
\IEEEauthorrefmark{1}Green Canyon High School, 2960 Wolf Pack Wy North, North Logan, UT USA. \\
\IEEEauthorrefmark{2}School of Electrical and Computer Engineering, University of Georgia, Athens, GA  USA. \\
\IEEEauthorrefmark{3}Department of Electrical and Computer Engineering, University of Utah, Salt Lake City, UT  USA. \\
Emails: \IEEEauthorrefmark{1}ethanzhu987@gmail.com, \IEEEauthorrefmark{2}hsun@uga.edu,
\IEEEauthorrefmark{3}mingyue.ji@utah.edu
}}
\maketitle

\begin{abstract}
Connected and automated vehicles (CAVs) have become a transformative technology that can change our daily life.
Currently, millimeter-wave (mmWave) bands are identified as the promising CAV connectivity solution. While it can provide high data rate, their realization faces many challenges such as high attenuation during mmWave signal propagation and mobility management. Existing solution has to initiate pilot signal to measure channel information, then apply signal processing to calculate the best narrow beam towards the receiver end to guarantee sufficient signal power. This process takes significant overhead and time, hence not suitable for CAVs. In this study, we propose an autonomous and low-cost testbed to collect extensive co-located mmWave signal and other sensor data such as LiDAR (Light Detection and Ranging), cameras, ultrasonic, etc, traditionally for ``automated'', to facilitate mmWave vehicular communications. Intuitively, these sensors can build a 3D map around the vehicle and signal propagation path can be estimated, eliminating the iterative  process via pilot signals. This multimodal data fusion, together with AI, is expected to bring significant advances in “connected” research. 
\end{abstract}


\section{Introduction}

More than 40,000 people die each year on U.S. highways, 1.3 million worldwide \cite{WinNT}. Efforts from all aspects are needed to reduce such a high fatality rate. Among them, CAV is the only solution that can bring the number to nearly 0, hence it has attracted increasing attentions from industry, academia, and government. As the name suggests, CAV involves two interconnected concepts: “connected” and “automated”. Recent years have witnessed significant advances in “automated” vehicles, creating self-driving capability up to Level 5 (fully autonomous). Current autonomous control is implemented in a more isolated way, made possible by sensors, pre-trained AI algorithms and on-board computer processing within individual vehicle. With more testing and deployed CAVs in the near future, we expect there is a pressing need to provide capability to exchange, transmit, and collect real-time data via vehicle-to-vehicle (V2V), vehicle-to-infrastructure (V2I), vehicle-to-everything (V2X) communications, i.e., the need for “connected”.

Dedicated short range communication (DSRC) is the first standard for vehicle communication that operates on 5.9 GHz and delivers data rate from 3 to 27 Mb/s \cite{kenney2011dedicated}. But it soon encounters bottleneck as V2V and V2I desire higher data rate and lower latency for assisted driving. Moreover, DSRC requires the installation of many new devices at the roadside, which becomes a barrier for wide deployment. State-of-the-art proposes to utilize 5G cellular millimeter-wave (mmWave) band ($\sim$28 GHz and above), an evolution from current cellular V2X (C-V2X), to address these challenges \cite{chen2023cellular}. As a core innovation for 5G, mmWave will be readily available with future base station (BS) infrastructures. However, applying it for vehicle scenario is still in the infancy and faces many unique challenges \cite{zhang2023map2schedule}. The main reason is that mmWave requires directional transmission to compensate for high signal attenuation, but it needs to send pilot signals to obtain precise propagation information, then calculates the best direction. This iterative process takes too long time and therefore is not suitable for high mobility scenario, where timely alerts, warning, and advice on driving decisions are critical. 

Alternatively, some works have proposed a sensor fusion approach to align mmWave transmissions in nearly real-time. Specifically, this approach is to utilize ``out-of-band'' information from LiDAR, camera, ultrasonic sensors that are already mounted on CAVs to establish, maintain, and steer mmWave transmissions. Moreover, fueled by cutting-edge AI algorithms, we expect this approach can bring real-time capabilities. Briefly, LiDAR provides 3-D point cloud for surrounding objects; cameras give high-definition vision view; ultrasonic depicts pedestrians and objects in close proximity. These data, when combined, give “eyes” to automated vehicles. On the other hand, signal propagation follows fundamental physics law that reflects, diffracts, or diffuses when encountering objects, hence its behavior is statistically predictable when precise perception is available. In fact, similar to video games, wireless simulators use “ray tracing” to model all possible signal paths. It is therefore a natural choice to combine sensor data and vehicle communications. 

Along this direction, to date, a few works have explored to using the out-of-band information for facilitating mmWave transmission. For example, in \cite{klautau2019lidar}, the authors used cameras to localize receiver and then guide the transmitter to use direct beamforming towards the receiver. This approach, however, will not likely work well for non-line-of-sight (NLoS) scenarios. Then in \cite{xu20203d}, the proposed multi-sensor fusion that consists of input from radar and camera  has shown an improved performance even under NLoS. In \cite{alrabeiah2020viwi}, a field measurement dataset is constructed, which contains LiDAR, camera, radar, and mmWave signal directions. The tested scenario is also versatile, from vehicle (dynamic) to pedestrians (relative static). However, existing testbeds suffer from high cost and are difficult to configure and use, which limits their potentials.

To summarize, this paper has the following contributions. 

\begin{itemize}
    \item The proposed HawkRover consists of commercial-off-the-shelf (COTS) TP-Link Talon AD7200 routers and a RC car. This provides a cost-effective solution for obtaining mmWave wireless and co-located multi-sensory data. All the data are time-synchronized through Internet time. 
    \item A deep learning (DL) algorithm is applied to predict the mmWave beamforming direction from sensory input, eliminating the signal-based pilot estimation approach.  
    \item Our results show that the constructed dataset and the DL algorithm perform well in real-world data, proving their potential in future V2X mmWave networks.  
\end{itemize}

The rest of the paper is organized as follows. Section II introduces the system architect design, followed by the sensor-aided mmWave beamforming alignment dataset and algorithm. We give a case-study in Section IV and conclude this paper in Section V. 

\section{System Architect Design and Data Collection}

\subsection{System overview and implementation}
\begin{figure}[ht]
	\includegraphics[width=3.3in]{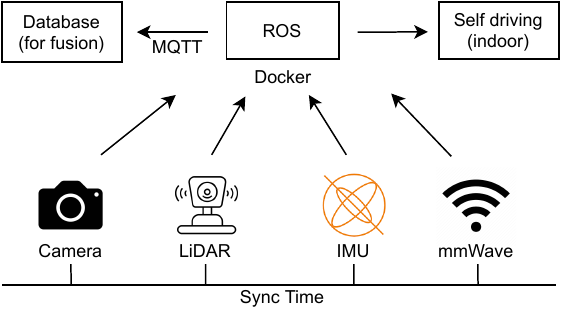}
	\centering
	\caption{HawkRover system architect}
	\label{fig:system}
	\centering
\end{figure}

The HawkRover system architect is shown in Fig. \ref{fig:system}. This overall design is versatile, allowing for adding sensors in a plug-and-play fashion. In particular, we use ROS (Robot Operating System) as the main software platform. ROS is a middleware between real operating system (Linux) and hardware. It contains a set of software libraries that are optimized to interact with sensor devices and has been widely adopted in robot community \cite{quigley2009ros}. Docker is a virtualization suite and can dynamically allocate ROS system resources to each sensor, hence allowing for better utilization on limited onboard resources. ROS will collect data from cameras, LiDAR,  inertial measurement unit (IMU), and mmWave modules. When generating data, all sensors retrieve global synchronized time, which is critical for fusion algorithm. For indoor tests, data from cameras, LiDAR and IMU will also be used to enable autonomous driving on RC car, so that this platform can operate without intervention. Lastly, since a huge amount of data will be generated, an external storage unit is deployed through wireless connection. Efficient data transmission between ROS and storage unit is enabled by message queuing telemetry transport (MQTT), a lightweight, loss-less, and bi-directional network layer protocol. Below is a brief list of main tools used in this HawkRover.

\emph{Software:}  ROS-Melodic on Linux, Python, Docker, MQTT, sensor APIs. 

\emph{Hardware:} Remote control (RC) car as the indoor vehicle, Nvidia Jetson (main onboard computing device), several wide-angle cameras, Slamtec RPLiDAR as the LiDAR, Adafruit IMU sensor, commercial mmWave router from TP-Link Talon AD7200 series with 32 analog beamforming antennas, and several battery packs to power each component. 

ROS allows the \emph{publisher/subscriber} mode,  where each sensor interface acts as a publisher to push the generated data to the subscriber (Linux).  A timestamp is also recorded when data are admitted. The data will be temporarily stored on an onboard device before entering the MQTT pipeline, which connects to a much larger external storage unit. 
The mmWave device we choose is not only cost effective and power efficient, but also has a simple beamforming mechanism (pre-defined codebook and exhaustive beam sweep), which is likely to be adopted in future vehicular communication. A complete set of indoor package costs around \$1,300. This is possible with the low-cost mmWave router and RC car. Our setup is much lower than existing solutions. For example, one pair of mmWave TX and RX unit from universal software radio peripheral (USRP), each equipped with a flexible number of antennas, can easily cost more than \$200,000 \cite{zhang2022testbed}.  The cost of \cite{alrabeiah2020viwi} can exceed \$10,000 due to their higher-end mmWave devices. 

\begin{figure}[ht]
    \centering
    \includegraphics[width=0.79\linewidth]{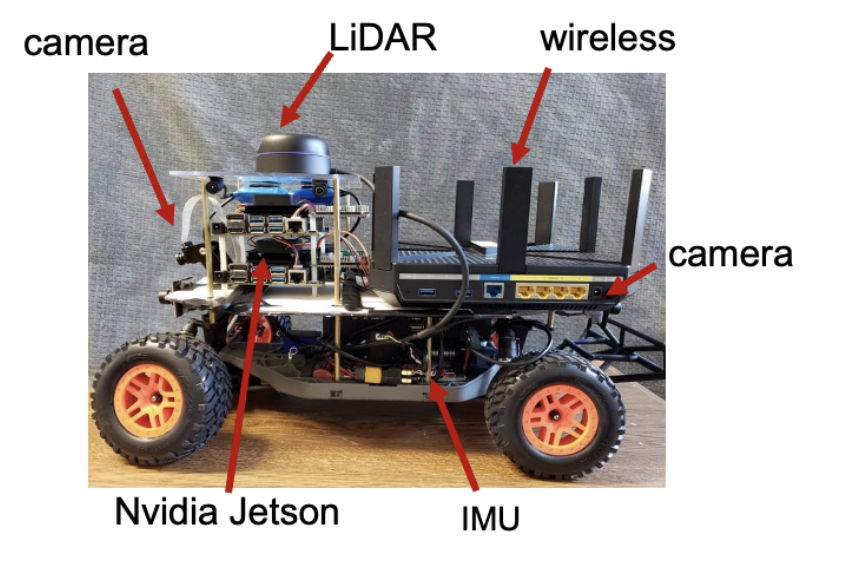}
    \caption{HawkRover data collection platform}
    \label{rover}
\end{figure}

As mentioned before, the proposed testbed can be operated autonomously, thanks to the ``self-driving'' capability enabled by the Donkey car \cite{Donkey}. It utilizes the front camera to identify the edges of the road and continuously control the steering angles and throttles. In the lab, we have setup a large canvas with dark tapes acting as the ``road''. Fig. \ref{rover} shows the prototype of HawkRover.

\subsection{Lab data  collection}
\begin{figure}[h]
    \centering
    \includegraphics[width=0.8\linewidth]{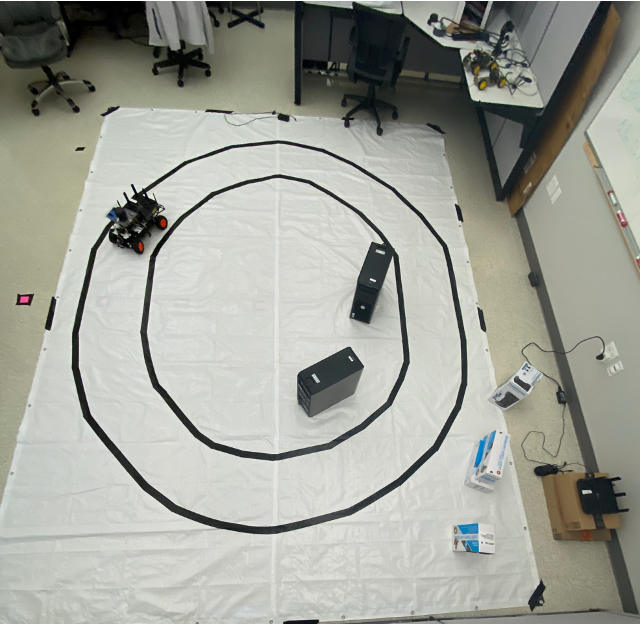}
    \caption{Lab data collection campaign}
    \label{Lab}
\end{figure}

Following pre-defined trajectory, HawkRover can run up to 3 hours without intervention, and each run session can collect up to 45 Gb of all sensory data. The cap of the run time is due to battery limit. As shown in Fig. \ref{Lab}, we place a large canvas in different locations of the lab room that has random objects in the scene, including office cubic, chairs, desks. In the setup, HawkRover moves along the trajectory on the canvas while another Talon router (as the BS) is placed at the corner of the room. To create NLoS scenario, computer towers and some boxes are randomly placed between HawkRover and the BS.

TP-Link mmWave router is re-programmed with customized firmware \cite{steinmetzer2017compressive,pajovic2019fingerprinting,koike2020fingerprinting, wang2019fingerprinting}. With our script, HawkRover can sequentially scan a total of 36 pre-defined beamformers, which spans the entire spatial directions. For each beamformer, the BS will return a signal-to-noise (SNR) value, indicating its quality. Therefore, the transmitter side will use the beamformer with the largest SNR values. 
Due to firmware limitations, one complete scan takes around 100 ms, resulting a scan frequency of 10 Hz. Both front and rear cameras are able to record images at 30 Hz, each with a resolution of $1280\times720$. The LiDAR can send 1,600 laser beams at 15 Hz, each laser can be translated into 3D point cloud. The IMU sensor is used to record HawkRover relative location and speed information, the frequency is adjustable. In this setup, we collect data at 100 Hz, each data point consists of acceleration, magnetic field, and orientation from 3 dimension $(x,y,z)$.   Table .\ref{table1} summarizes all data and their format. 

\begin{table} [h]
\small
\centering
\begin{tabular}{| c| c | c |}

    \hline
    Sensor & Data frequency (Hz) & Data \# per sample \\
    \hline
    camera & 30  & $1280\times 720$ \\
    \hline
    LiDAR &  15  &   $ 1600 \times (x,y,z)$ \\
    \hline 
    IMU                    & 100      & 10 \\ 
    \hline
    mmWave sensor & $ \sim$10 & 36 \\
    \hline
     location  & 100  & 2 \\
    \hline
\end{tabular}
\caption{Data format from different sensors on HawkRover} \label{table1}
\end{table}

\section{Sensor-aided mmWave Beam Alignment}
In the previous section, we have introduced the HawkRover platform and the data format. With HawkRover, enormous  amount of data and be collected in a cost-effective and flexible way. In the following, we focus on the data processing and DL algorithm.  
\subsection{Dataset pre-processing}
\begin{figure}[h]
    \centering
    \includegraphics[width=1.0\linewidth]{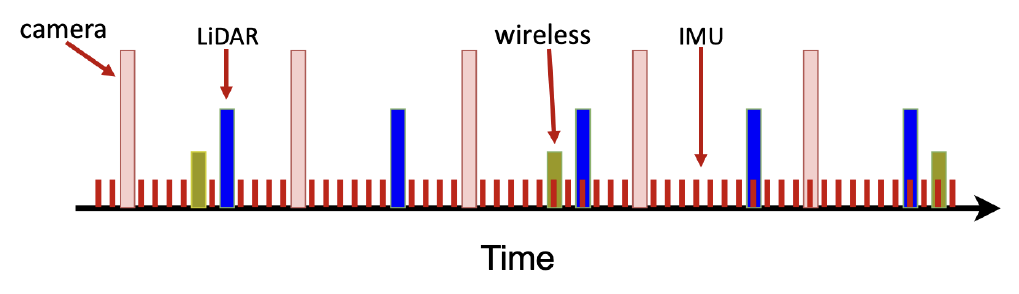}
    \caption{Collected time-series data}
    \label{asy}
\end{figure}

From Table \ref{table1}, one can observe that although all sensor data are synchronized, their events (the time of recording) are not aligned in the time domain due to different data frequencies. This phenomenon is also illustrated in Fig. \ref{asy}.  In fact, mmWave sensor data is very spare, which poses challenges to create the mapping relationship between beam directions and propagation environment sensing input. As an example, consider the mmWave sensor generated one sample data, while camera and LiDAR may have produced more than three events that contains excessive information that related to the mmWave signal. 

Therefore, the system may suffer over-fitting issue from vision sensors when predicting mmWave communication beamformer. There are many algorithms to handle this challenges, mostly from autonomous driving field, such as in \cite{dalgaty2023hugnet}. Besides, algorithms with  long short-term memory (LSTM) is often utilized to process time-series context information. To reduce the complexity, we use another simplified approach. Specifically, the following steps are proceed: 1) the mmWave sensor data is kept, together with its timestamp. 2) Smooth out camera and LiDAR data with point-wise interpolation, and then only keep the data that is closest to mmWave sensor timestamp. 3) Normalize mmWave SNR data since it has a high dynamic range. 4) Process IMU data and translate into relative positions. 

Although the pre-processing techniques used here can significantly reduce data amount (by removing excessive vision data), one drawback is the context information may be lost. On the other hand, the advantage is the much lower overhead for data input. 

\subsection{Beam prediction with multimodal sensor fusion}

The aim is to directly predict the best pair of codebooks from multisensor input, skipping channel estimation, as well as exhaustive search stage.  Since enormous datasets can be generated from the HawkRover platform, our specific goal is to directly find the mapping:
\begin{equation}
    (b_{UE}^*, b_{BS}^*) = \max_{b_{UE}, b_{BS}} f (\mathcal{C, L, I, P}),
\end{equation}
where $ (b_{UE}^*, b_{BS}^*)$ is the optimal beam codebook index pair (same as mmWave direction). $\mathcal{C}, \mathcal{L}, \mathcal{I}, \mathcal{P}$ are camera, LiDAR, IMU, and position data, respectively. $f$ is the mapping function to be developed through data-driven DL.

\begin{figure}[h]
    \centering
    \includegraphics[width=0.9\linewidth]{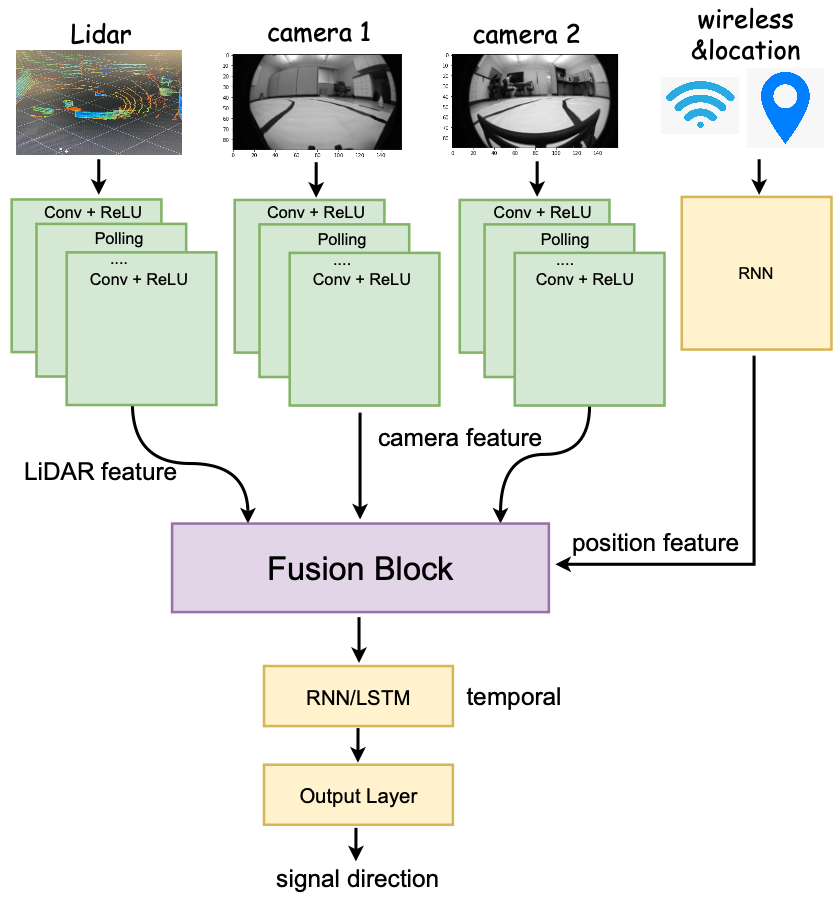}
    \caption{Multi-sensory data fusion with DL}
    \label{fig:enter-label}
\end{figure}

In Fig. \ref{fig:enter-label}, the overall structure of the fusion algorithm is presented. At the high level, each sensor's input is processed through convolutional neural network (CNN) and recurrent neural network (RNN). The former is usually used to process image, while the latter is for sequential data, perfectly fit for numerical wireless and location data. A fusion block is applied to concatenate each extracted feature, and then all features are fed into LSTM block for temporal correlations. Lastly, the softmax-based output layer is utilized to determine the beam index with highest probability. 

In CNN, convolution (Conv), rectified linear unit (ReLU), and pooling are three fundamental building blocks. Each serves a distinct function. Conv layers perform a mathematical operation that takes two inputs, such as an image matrix and a filter or kernel, and produces a feature map. This operation involves sliding the filter over the image and computing the dot product at each position. Convolution preserves the spatial relationship between pixels by learning image features based on small squares of input data. The ReLU function applies a non-linear activation to the output of the convolution operation. It replaces all negative pixel values in the feature map with zero. This non-linearity introduces the ability to solve more complex problems by allowing the network to create complex mappings from the input to the output, improving the network's ability to learn from the data.  Pooling reduces the dimensionality of each feature map while retaining the most important information. Max pooling, for example, slides a window over the input and takes the maximum value at each step. By doing so, pooling layers reduce the number of parameters, lessen computational complexity, and provide a form of translation invariance. They help the network to be more robust to the exact location of features in the input space.

For RNN, it has the ability to maintain a form of memory by using their internal state (hidden layers) to process sequences of inputs. RNNs can process inputs of varying lengths because they process data sequentially and iteratively. This makes them well-suited for tasks like time series prediction. Besides, RNNs have a hidden state that captures information about what has been calculated so far. In essence, the hidden state serves as the network’s memory, which contains information about previous inputs, allowing the network to make decisions based on historical context.
This is achieved by having loops in the network, allowing wireless and location information to persist. 

Lastly, the reason for feature fusion, instead of direct data fusion, is due to the vision data high dimensionality.  

\section{A Case Study and Results}
In this section, we provide a case study for one particular dataset collected. The location is a typical lab space. Note that during data collection, normal daily activities are not restricted, for example, people walking in the room, lab furniture movements, etc. This, in turn, shows the robustness and realism of the dataset. 

\begin{figure}[h]
    \centering
    \includegraphics[width=0.7\linewidth]{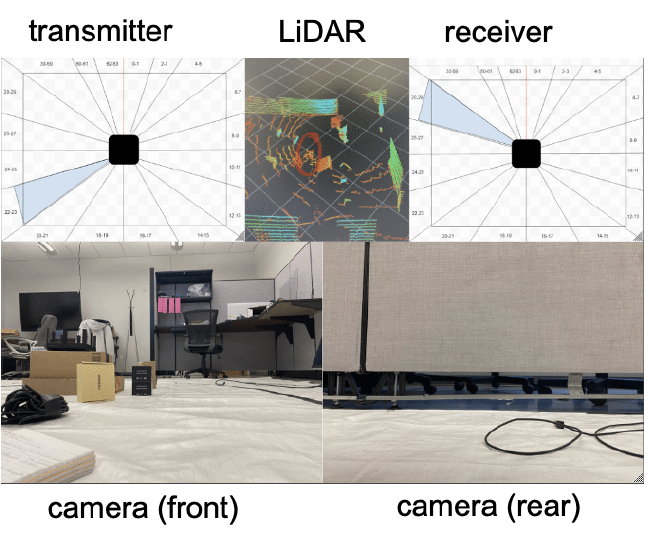}
    \caption{Visualization of the collected data}
    \label{fig:sample}
\end{figure}

In Fig. \ref{fig:sample}, the collected sample is visualized. Specifically, we show the beamforming directions coming from both the transmitter (HawkRover) side and the BS side. Besides, the vision data from LiDAR and two camera are also illustrated simultaneously. These data are perfectly synchronized through network time protocol (NTP) through high-speed local network, with error no more than 1 ms, which is negligible compared with the sampling rate from these sensors. 

\begin{figure}
    \centering
\includegraphics[width=1\linewidth]{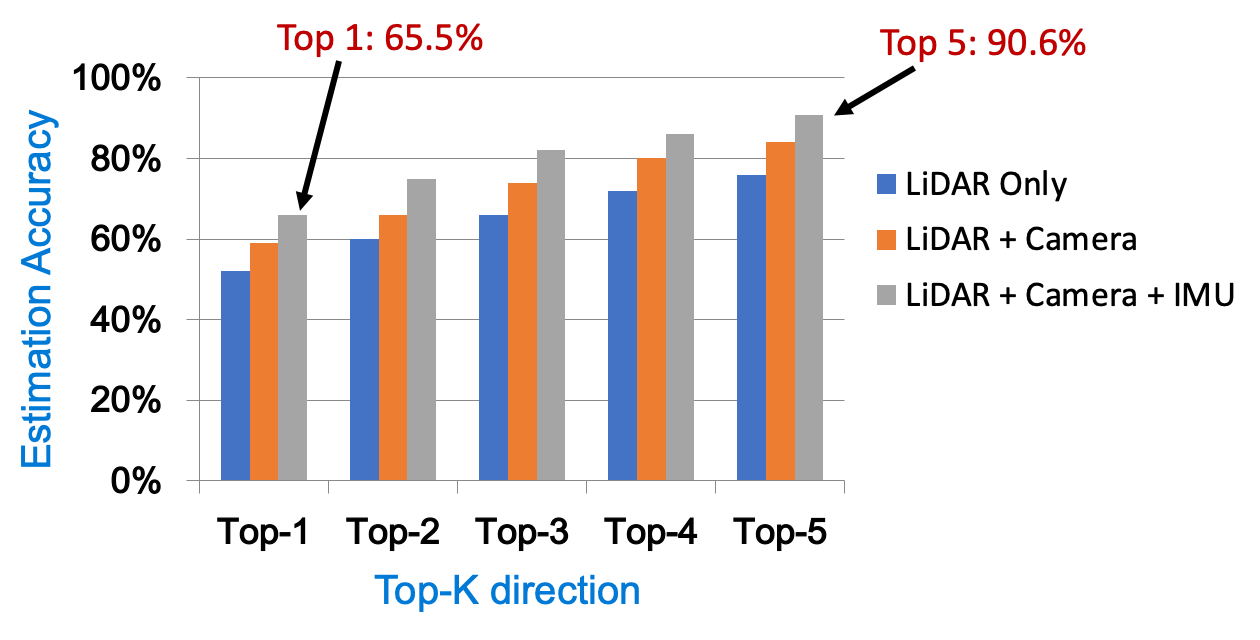}
    \caption{mmWave beamforming prediction results}
    \label{result}
\end{figure}

The collected data is divided into two parts: training (80\%) and testing (20\%). Both training process and testing experiments are implemented in PyTorch. 
As shown in Fig. \ref{result}, the optimal beamforming prediction result based on sensor fusion algorithm is presented. In DL, Top-$K$ is often used as an evaluation metric to assess the performance. For example, in our case, top-1 accuracy means the model's most probable prediction must be correct for it to count as a success. Meanwhile, top-5 accuracy means any of the model's five most probable predictions can be correct. We showcase the results from top-1 to top-5, each with three different settings, namely, LiDAR only, LiDAR + camera fusion, and LiDAR + Camera + IMU position fusion.  

Top-1 with LiDAR, camera, and IMU fusion provides prediction accuracy of 65.5\%, which is comparable to results in \cite{jiang2022computer}. Note that in their work, line-of-sight (LoS) is dominated and higher-end devices (both mmWave and sensing) are used. Besides, it is readily shown that prediction accuracy increases with higher $K$ values, reaching to 90.6\% in top-5. This performance gives confidence of our proposed fusion algorithm, which suggests that DL can accurately predict beam direction from ``out-of-band'' information, skipping the process of signal-based pilots. Even with only LiDAR, top-5 is close to 80\%, highlighting the strong correlation of propagation environments and the mmWave spatial characteristics. 

We need to note that the mmWave spatial fingerprints of the lab room may have also contributed to this performance. As shown in our early works \cite{pajovic2019fingerprinting, wang2019fingerprinting, koike2020fingerprinting}, mmWave signal can have a spatial resolution to sub-10 cm, and the DL may have learnt this implicit relationship of location and optimal beam directions. Nevertheless, sensor fusion approach can significantly reduce the beam tracking overhead, making it highly desirable feature for vehicular scenarios.

\section{Conclusions}
mmWave bands are emerging as a viable solution for CAV communication. However, their high attenuation and mobility management pose considerable challenges. Our study introduces HawkRover, an autonomous and cost-effective testbed for collecting comprehensive mmWave and sensor data such as from LiDAR, cameras, and IMUs, to enhance vehicular communication. By using DL-based sensor fusion, our method can predict signal paths more efficiently than iterative pilot signal processes. A case study for indoor scenario is given, which promises substantial progress in CAV communication research.

\section{Acknowledgement}
The work of H. Sun is partially supported by The University of Georgia E-Mobility Initiative seed grant. J. Schnor and S. Flynn also provided support for the prototype development. 
\bibliographystyle{IEEEtran}
\bibliography{lib}

\begin{thebibliography}{10}
\providecommand{\url}[1]{#1}
\csname url@samestyle\endcsname
\providecommand{\newblock}{\relax}
\providecommand{\bibinfo}[2]{#2}
\providecommand{\BIBentrySTDinterwordspacing}{\spaceskip=0pt\relax}
\providecommand{\BIBentryALTinterwordstretchfactor}{4}
\providecommand{\BIBentryALTinterwordspacing}{\spaceskip=\fontdimen2\font plus
\BIBentryALTinterwordstretchfactor\fontdimen3\font minus
  \fontdimen4\font\relax}
\providecommand{\BIBforeignlanguage}[2]{{%
\expandafter\ifx\csname l@#1\endcsname\relax
\typeout{** WARNING: IEEEtran.bst: No hyphenation pattern has been}%
\typeout{** loaded for the language `#1'. Using the pattern for}%
\typeout{** the default language instead.}%
\else
\language=\csname l@#1\endcsname
\fi
#2}}
\providecommand{\BIBdecl}{\relax}
\BIBdecl

\bibitem{WinNT}
``{World Health Organization (WHO)} global status report on road safety 2018,''
  \url{https://www.who.int/publications/i/item/9789241565684}, accessed:
  2023-10-1.

\bibitem{kenney2011dedicated}
J.~B. Kenney, ``Dedicated short-range communications (dsrc) standards in the
  united states,'' \emph{Proceedings of the IEEE}, vol.~99, no.~7, pp.
  1162--1182, 2011.

\bibitem{chen2023cellular}
S.~Chen, J.~Hu, L.~Zhao, R.~Zhao, J.~Fang, Y.~Shi, and H.~Xu, \emph{Cellular
  Vehicle-to-Everything (C-V2X)}.\hskip 1em plus 0.5em minus 0.4em\relax
  Springer Nature, 2023.

\bibitem{zhang2023map2schedule}
L.~Zhang, H.~Sun, J.~Sun, R.~Parasuraman, Y.~Ye, and R.~Q. Hu, ``Map2schedule:
  An end-to-end link scheduling method for urban v2v communications,'' 2023.

\bibitem{klautau2019lidar}
A.~Klautau, N.~Gonz{\'a}lez-Prelcic, and R.~W. Heath, ``Lidar data for deep
  learning-based mmwave beam-selection,'' \emph{IEEE Wireless Communications
  Letters}, vol.~8, no.~3, pp. 909--912, 2019.

\bibitem{xu20203d}
W.~Xu, F.~Gao, S.~Jin, and A.~Alkhateeb, ``3d scene-based beam selection for
  mmwave communications,'' \emph{IEEE Wireless Communications Letters}, vol.~9,
  no.~11, pp. 1850--1854, 2020.

\bibitem{alrabeiah2020viwi}
M.~Alrabeiah, A.~Hredzak, Z.~Liu, and A.~Alkhateeb, ``Viwi: A deep learning
  dataset framework for vision-aided wireless communications,'' in \emph{2020
  IEEE 91st Vehicular Technology Conference (VTC2020-Spring)}.\hskip 1em plus
  0.5em minus 0.4em\relax IEEE, 2020, pp. 1--5.

\bibitem{quigley2009ros}
M.~Quigley, K.~Conley, B.~Gerkey, J.~Faust, T.~Foote, J.~Leibs, R.~Wheeler,
  A.~Y. Ng \emph{et~al.}, ``Ros: an open-source robot operating system,'' in
  \emph{ICRA workshop on open source software}, vol.~3, no. 3.2.\hskip 1em plus
  0.5em minus 0.4em\relax Kobe, Japan, 2009, p.~5.

\bibitem{zhang2022testbed}
Q.~Zhang and C.~Yang, ``Testbed and performance evaluation of 3d mmwave beam
  tracking in mobility scenario,'' in \emph{IEEE INFOCOM 2022-IEEE Conference
  on Computer Communications Workshops (INFOCOM WKSHPS)}.\hskip 1em plus 0.5em
  minus 0.4em\relax IEEE, 2022, pp. 1--2.

\bibitem{Donkey}
``An opensource diy self driving platform for small scale cars.''
  \url{https://www.donkeycar.com}, accessed: 2023-10-1.

\bibitem{steinmetzer2017compressive}
D.~Steinmetzer, D.~Wegemer, M.~Schulz, J.~Widmer, and M.~Hollick, ``Compressive
  millimeter-wave sector selection in off-the-shelf ieee 802.11 ad devices,''
  in \emph{Proceedings of the 13th International Conference on emerging
  Networking EXperiments and Technologies}, 2017, pp. 414--425.

\bibitem{pajovic2019fingerprinting}
M.~Pajovic, P.~Wang, T.~Koike-Akino, H.~Sun, and P.~V. Orlik,
  ``Fingerprinting-based indoor localization with commercial mmwave wifi-part
  i: Rss and beam indices,'' in \emph{2019 IEEE Global Communications
  Conference (GLOBECOM)}.\hskip 1em plus 0.5em minus 0.4em\relax IEEE, 2019,
  pp. 1--6.

\bibitem{koike2020fingerprinting}
T.~Koike-Akino, P.~Wang, M.~Pajovic, H.~Sun, and P.~V. Orlik,
  ``Fingerprinting-based indoor localization with commercial mmwave wifi: A
  deep learning approach,'' \emph{IEEE Access}, vol.~8, pp. 84\,879--84\,892,
  2020.

\bibitem{wang2019fingerprinting}
P.~Wang, M.~Pajovic, T.~Koike-Akino, H.~Sun, and P.~V. Orlik,
  ``Fingerprinting-based indoor localization with commercial mmwave wifi-part
  ii: Spatial beam snrs,'' in \emph{2019 IEEE Global Communications Conference
  (GLOBECOM)}.\hskip 1em plus 0.5em minus 0.4em\relax IEEE, 2019, pp. 1--6.

\bibitem{dalgaty2023hugnet}
T.~Dalgaty, T.~Mesquida, D.~Joubert, A.~Sironi, P.~Vivet, and C.~Posch,
  ``Hugnet: Hemi-spherical update graph neural network applied to low-latency
  event-based optical flow,'' in \emph{Proceedings of the IEEE/CVF Conference
  on Computer Vision and Pattern Recognition}, 2023, pp. 3952--3961.

\bibitem{jiang2022computer}
S.~Jiang and A.~Alkhateeb, ``Computer vision aided beam tracking in a
  real-world millimeter wave deployment,'' in \emph{2022 IEEE Globecom
  Workshops (GC Wkshps)}.\hskip 1em plus 0.5em minus 0.4em\relax IEEE, 2022,
  pp. 142--147.

\end{thebibliography}

\end{document}